\documentclass[%
preprint,
superscriptaddress,
 showkeys,
 amsmath,amssymb,
 aps,
]{revtex4-1}

\usepackage{graphicx}% Include figure files
\usepackage{dcolumn}% Align table columns on decimal point
\usepackage{bm}% bold math
\usepackage{hyperref}% add hypertext capabilities
\usepackage{epstopdf}

	\usepackage{fancyheadings}
\renewcommand{\thispagestyle}[1]{} % do nothing

\begin{document}

%\preprint{APS/123-QED}

\title{THERMODYNAMIC PROPERTIES OF A HUBBARD MODEL\\ON A CUBIC CLUSTER -- EXACT DIAGONALIZATION STUDY AT QUARTER FILLING}

\author{Karol Sza\l{}owski}
 \email{kszalowski@uni.lodz.pl}
\author{Tadeusz Balcerzak}%
\affiliation{%
 Department of Solid State Physics, Faculty of Physics and Applied Informatics, University of \L{}\'od\'z, ul. Pomorska 149/153, 90-236 \L{}\'od\'z, Poland
}%

\author{Michal Ja\v{s}\v{c}ur}
\author{Andrej Bob\'{a}k}
\author{Milan \v{Z}ukovi\v{c}}
\affiliation{
Department of Theoretical Physics and Astrophysics, Faculty of Science, \\P. J. \v{S}\'{a}f\'{a}rik University, Park Angelinum 9, 041 54 Ko\v{s}ice, Slovak Republic
}%


\begin{abstract}
We study the thermodynamics of a zero-dimensional, cubic cluster described with a Hubbard Hamiltonian, focusing our interest on the magnetic properties. The range in which the studied cluster is paramagnetic is considered. The results are obtained by means of exact numerical diagonalization. Such thermodynamic quantities as entropy, specific heat, magnetic susceptibility, spin-spin correlations and double occupancy are discussed. Particular emphasis is put on the behaviour of local maxima of specific heat and susceptibility, which are analysed in terms of Schottky anomalies. 
\end{abstract}

\maketitle

%\tableofcontents

\vspace{3mm}

\section{Introduction}

Low-dimensional magnetic systems attract considerable theoretical and experimental efforts. Within this field, noticeable attention is paid to the theoretical studies of zero-dimensional magnetic clusters composed of a finite, small number of atoms \cite{Zukovic2014,Strecka2014,Zukovic2015,Strecka2015}. Although the existence of magnetic ordering and magnetic phase transitions in low dimensions is severely limited, yet such systems can still exhibit a range of interesting properties. 

Hubbard model \cite{Hubbard,Tasaki,Spalek2015} is one of the successful theoretical approaches involved, among others, in the studies of low-dimensional magnetic systems exhibiting strong correlations (to mention for example its early applications to magnetic thin films \cite{Wojtczak1,Wojtczak2,Wojtczak3}). In spite of its simplicity, it still remains a challenge for theorists. In addition to various approximate approaches to its thermodynamics, a valuable method for this model is exact diagonalization \cite{Feshke}, being essentially free from any artefacts, which feature is of huge importance due to the complicated phase diagram of the model. However, such method is computationally very demanding, therefore its applicability is limited only to the smallest systems. This feature allows to focus the interest on zero-dimensional clusters described by Hubbard model, the magnetic properties of which were studied for various geometries and numbers of charge carriers (filling levels of energy states) \cite{Callaway1986,Callaway1987,Callaway1988,Parola,Tan1991,Macedo,Lopez2009,Zwicker,Zwicker2}. In particular, cubic cluster attracted some attention \cite{Callaway1987,Zwicker,Zwicker2}. However, not all the thermodynamic properties related to magnetism were systematically explored, especially beyond the half-filling case. 

The aim of the present paper is to provide a systematic discussion of thermodynamics of a cubic Hubbard cluster, selecting the case of quarter-filling of the energy states.

\begin{figure}[h]
\begin{center}
\includegraphics[width=88mm]{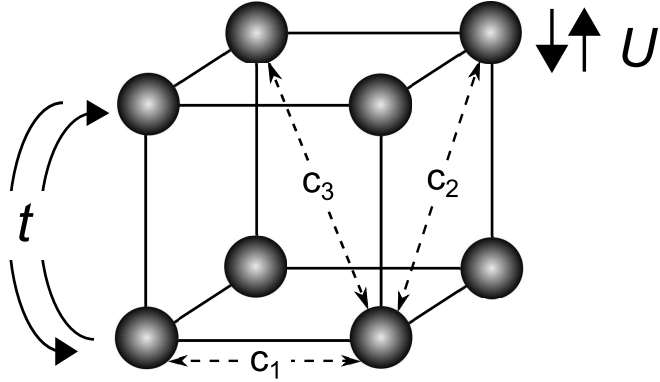}

\vspace{2mm}

\caption{A schematic view of a cubic cluster system described with a Hubbard Hamiltonian with the hopping integral $t$ and on-site Coulombic energy $U$. The spin-spin correlations between first, second and third nearest neighbours are indicated with dashed lines.\label{rys1}}
\end{center}
\end{figure}

\section{Theoretical model} 
 
A schematic view of the system of interest is shown in Fig.\,\ref{rys1}. It is a zero-dimensional cluster in a form of a cube, having $N=8$ sites with hopping integral $t$ between nearest neighbours only. The on-site Coulombic interaction energy is parametrized by the parameter $U>0$. In our study we focus on the case with $N_e=4$ charge carriers (electrons) in the system, thus we deal with quarter-filling of the energy states. Let us remind that such a molecular-like structure possesses a purely discrete energy spectrum, so we do not refer to band filling.  

The system is described by the following Hubbard Hamiltonian:
\begin{equation}
\mathcal{H}=-t\sum_{\left\langle i,j\right\rangle}^{}{\sum_{\sigma=\uparrow,\downarrow}^{}{\left(c^{\dagger}_{i,\sigma}c_{j,\sigma}+c^{\dagger}_{j,\sigma}c_{i,\sigma}\right)}}+U\sum_{i}^{}{n_{i,\uparrow}n_{i,\downarrow}},
\label{eq1}
\end{equation}
in which $c^{\dagger}_{i,\sigma}$ ($c_{i,\sigma}$) creates (annihilates) an electron with spin $\sigma$ at site labelled with $i=1,\dots,N$, while $n_{i,\sigma}=c^{\dagger}_{i,\sigma}c_{i,\sigma}$ is the number of electrons with spin $\sigma$ at site $i$. Let us mention that $z$-component of the spin at site $i$ is therefore equal to $s_{i}^{z}=\left(n_{i,\uparrow}-n_{i,\downarrow}\right)/2$.

For the case of $N=8$ lattice sites with $N_e=4$ electrons, the Hilbert space for the studied system is spanned by $$N_{s}=\frac{\left(2N\right)!}{N_e ! \left(2N-N_e\right)!}=1820$$

\noindent basis vectors. The resulting matrix $(N_s \times N_s)$ of the Hamiltonian $\mathcal{H}$ for the given case can be exactly diagonalized numerically, what yields the eigenvalues and corresponding eigenvectors.

\begin{figure}[h]
\begin{center}
\includegraphics[width=120mm]{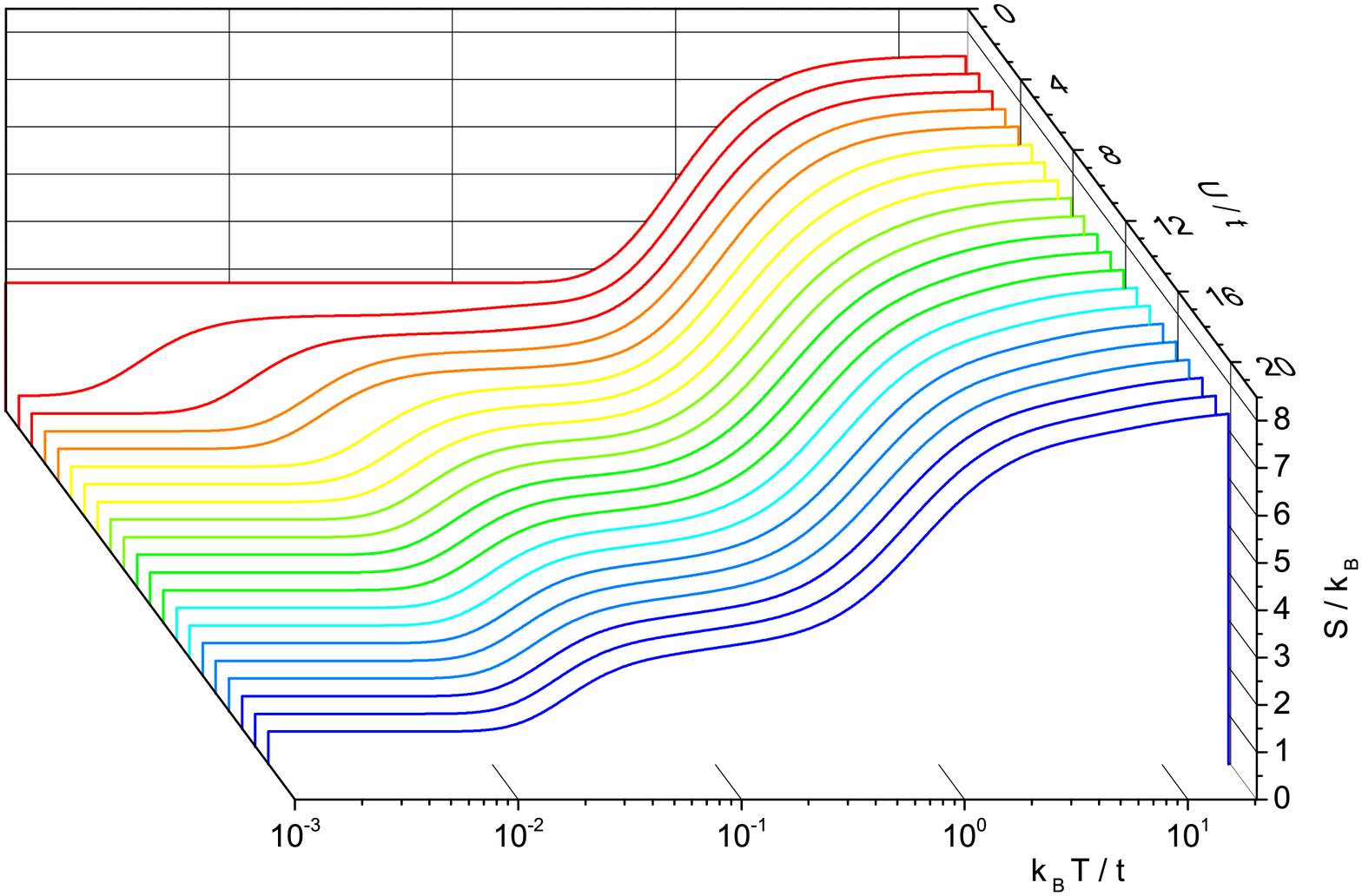}

\vspace{2mm}

\caption{The dependence of the system entropy on the normalized temperature for various values of normalized Coulombic interactions energy.\label{rys2}}
\end{center}
\end{figure}

The exact thermodynamic description of our system was constructed basing on the canonical ensemble, with fixed number of electrons and temperature equal to $T$. The statistical operator describing the thermal state is:
\begin{equation}
\rho = \frac{1}{\mathcal{Z}}\,e^{-\beta \mathcal{H}},
\end{equation}
where the statistical sum is given by:
\begin{equation}
\mathcal{Z}=\mathrm{Tr}\,e^{-\beta \mathcal{H}}=\sum_{k}{g_{k}\,e^{-\beta E_{k}}},
\end{equation}
$g_k$ being the degeneracy of the eigenstate with the energy equal to $E_k$, while $\beta=1/\left(k_{\rm B}T\right)$ with $k_{\rm B}$ denoting the Boltzmann constant. The thermodynamic average of an arbitrary quantity $A$ is equal to:
\begin{equation}
\left\langle A\right\rangle = \mathrm{Tr}\left(\rho A\right).
\end{equation}
Knowledge of the statistical operator and the statistical sum allows us to calculate all further thermodynamic quantities of interest, for instance, entropy, specific heat or magnetic susceptibility as well as magnetic correlations between first, second and third neighbours (see Fig.~\ref{rys1} for schematic explanation). In particular, specific heat $C_h$ and magnetic susceptibility $\chi_T$ can be determined conveniently using fluctuation-dissipation theorem, which yields:
\begin{equation}
C_{h}=k_{\rm B}\beta^2\left(\left\langle E^2\right\rangle-\left\langle E\right\rangle^2\right),
\label{eq3}
\end{equation}
where 
$$\left\langle E\right\rangle = \frac{1}{\mathcal{Z}} \sum\limits _{k} g_{k}E_{k}e^{-\beta E_{k}}
$$

\noindent is the internal energy 
and
\begin{equation}
\chi_{T}=\beta\left(\left\langle m^2\right\rangle-\left\langle m\right\rangle^2\right),
\label{eq2}
\end{equation}
where 
$$\left\langle m\right\rangle = \sum\limits _{i=1} ^{N} \left\langle s^{z}_{i} \right\rangle
$$

\noindent is the average total magnetization.

The entropy is calculated from $S=\left(\left\langle E\right\rangle-F\right)/T$, where $F=-k_{\rm B}T\ln \mathcal{Z}$ is Helmholtz free energy. 
The spin-spin correlations are defined as $c_k=\left\langle s^{z}_{i}s^{z}_{j} \right\rangle$, where $k=1,2,3$ means that spin at site $i$ is the $k$-th nearest neighbour of the spin at site $j$ (see Fig.~\ref{rys1}). Finally, the average double occupancy per site is expressed as 
$$d=\left(1/N\right)\displaystyle\sum_{i=1}^{N}{\left\langle n_{i,\uparrow}n_{i,\downarrow} \right\rangle}.$$

In our calculations we found that the ground state (at $T=0$) is nonmagnetic, with total spin equal to $S=0$ if $U/t\lesssim 223.7$, what is in agreement with the results of Refs.~\cite{Callaway1987,Zwicker} (while for stronger Coulombic interactions $U$ we deal with a ferromagnetic ground state). Since the critical strength of Coulombic interactions corresponds to extremely high $U/t$ values, we limit our considerations to the range with zero total ground state spin, assuming $0\leq U/t\leq 50$. We believe this range is the most physically interesting and the thermodynamic properties of the cubic Hubbard cluster at quarter filling for this range were not systematically explored.

\vspace{3mm}

\section{Numerical results and discussion}

In this section we discuss the numerical calculations of such thermodynamic quantities as entropy, specific heat and magnetic susceptibility as well as spin correlation functions and double occupancy performed within canonical ensemble approach with exact numerical diagonalization of the Hubbard Hamiltonian. For the numerical calculations we used \emph{Wolfram Mathematica} software \cite{Mathematica}.

The dependence of total entropy of the cubic cluster on the temperature in presented in Fig.~\ref{rys2} in logarithmic temperature scale, in a wide range of normalized Coulombic energies $U/t$. For $U/t=0$ (i.e. for a pure tight-binding model) a ground state exhibits 15-fold degeneracy, therefore, the entropy at $T=0$ reaches a noticeable residual value of $k_{\rm B}\ln 15\simeq 2.708\,k_{\rm B}$. On the contrary, for $U/t>0$ (in our $U/t$ range of interest) the ground state is 2-fold degenerate, so that the residual entropy is reduced to $k_{\rm B}\ln 2\simeq 0.693\,k_{\rm B}$. The limiting, high-temperature entropy is in all cases  $k_{\rm B}\ln 1820\simeq 7.507\,k_{\rm B}$. In Fig.~\ref{rys2} it can be noticed that some low- and high-temperature ranges can be seen, in which entropy rises fast with the temperature, signalizing high values of the specific heat.

\begin{figure}[h]
\begin{center}
\includegraphics[width=120mm]{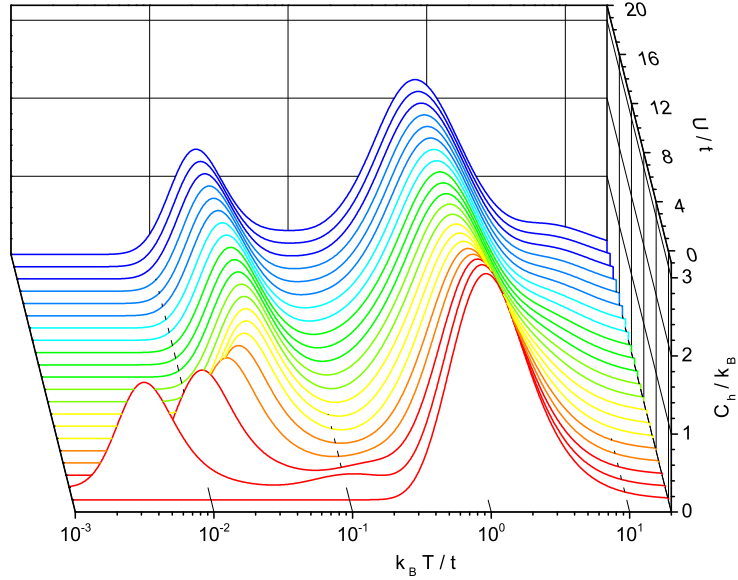}

\vspace{2mm}

\caption{The dependence of the system specific heat on the normalized temperature for various values of normalized Coulombic interactions energy.\label{rys3}}
\end{center}

\vspace{-3mm}

\end{figure}

The variation of specific heat with the temperature can be followed for a range of $U/t$ values in Fig.~\ref{rys3}. It is evident that in the absence of Coulombic interactions a single peak is present, at the temperature close to $k_{\rm B}T/t\simeq 1$, and it remains in that position when the Coulombic interaction $U/t$ is switched on. What is noticeable, appearance of $U/t>0$ causes a low-temperature specific heat maximum to emerge, as well as the significant shift of its position with the energy of Coulombic interactions, what can be clearly seen in Fig.~\ref{rys3}. 

\begin{figure}[h]
\begin{center}
\includegraphics[width=128mm]{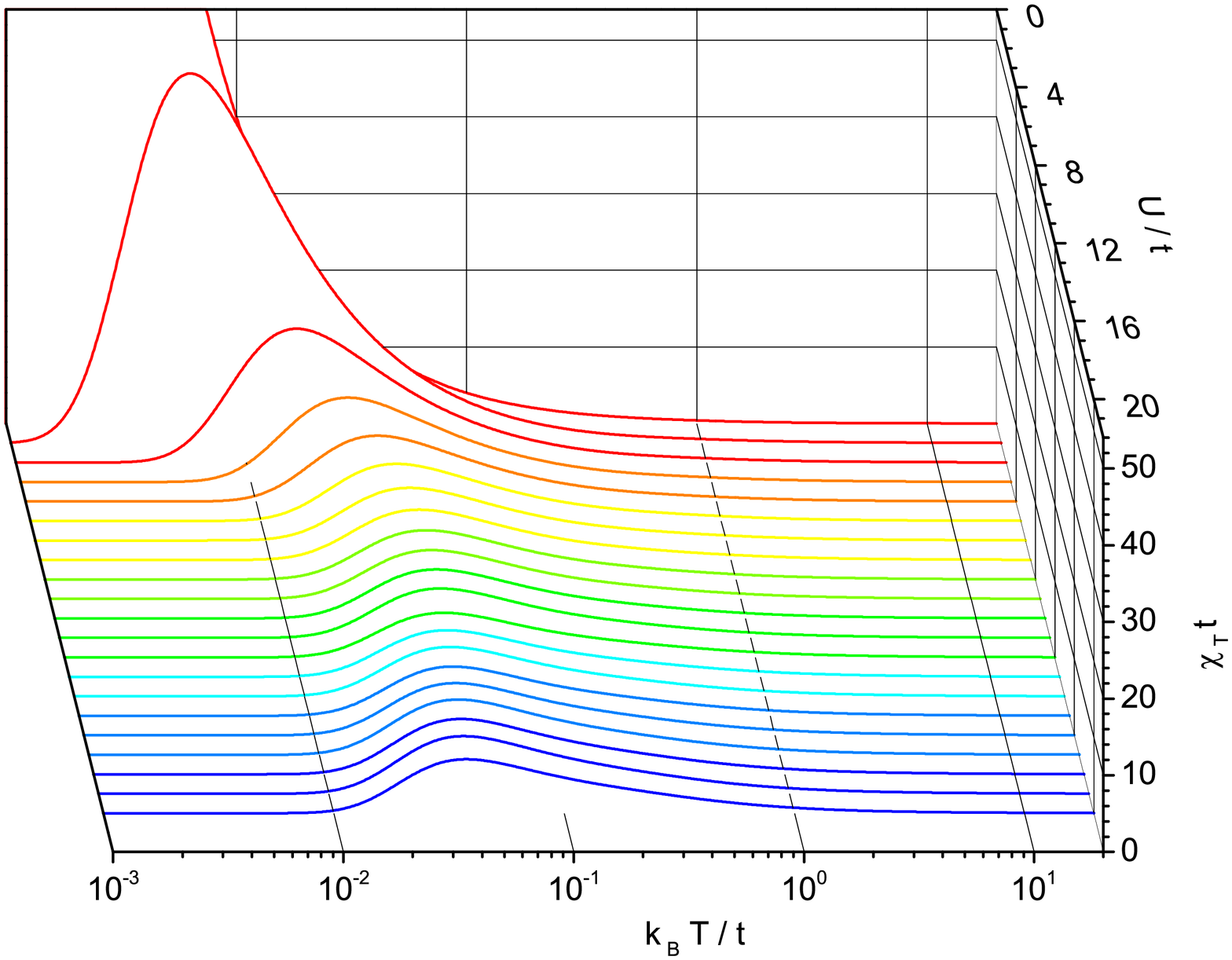}

\vspace{2mm}

\caption{The dependence of the system magnetic susceptibility on the normalized temperature for various values of normalized Coulombic interactions energy.\label{rys4}}
\end{center}
\end{figure}

Contrary to the specific heat, the magnetic susceptibility indicates only a single peak in its temperature dependence for $U/t>0$, as presented in Fig.~\ref{rys4}. The height of this maximum also significantly drops when Coulombic interactions become stronger, whereas both specific heat maxima show a rather constant height. It has been verified that the inverse of susceptibility shows a linear dependence on the temperature at high temperatures, what corresponds to Curie-Weiss law with positive (ferromagnetic) Curie-Weiss temperature (not shown in the plot); note that this fact does not imply the presence of magnetic ordering in our zero-dimensional system.

\begin{figure}[h]
\begin{center}
\includegraphics[width=125mm]{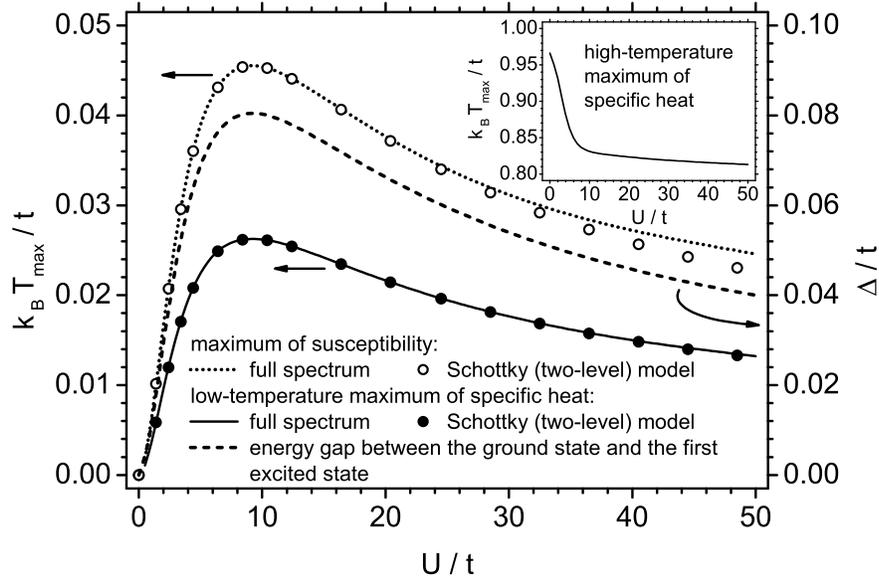}

%\vspace{2mm}

\caption{The dependence of the energy gap as well as of characteristic temparatures at which magnetic susceptibility and specific heat reach local maxima on the normalized Coulombic interactions energy.\label{rys5}}
\end{center}

\vspace{-3mm}

\end{figure}

\begin{figure}[h]
\begin{center}
\includegraphics[width=128mm]{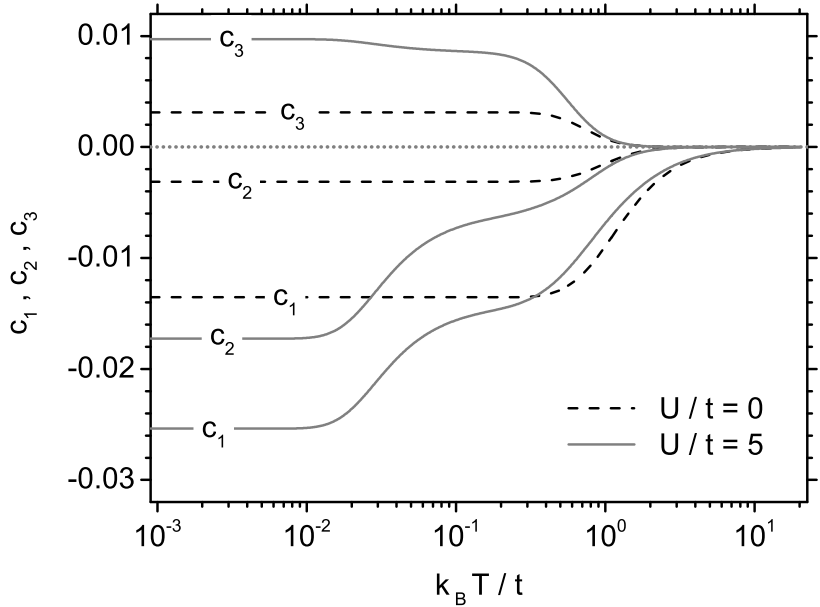}

\vspace{2mm}

\caption{The dependence of the spin-spin correlations for first, second and third neighbours (see Fig.~\ref{rys1}) on the normalized temperature for two representative values of normalized Coulombic interactions energy.\label{rys6}}
\end{center}

\vspace{-3mm}

\end{figure}

\begin{figure}[h]
\begin{center}
\includegraphics[width=128mm]{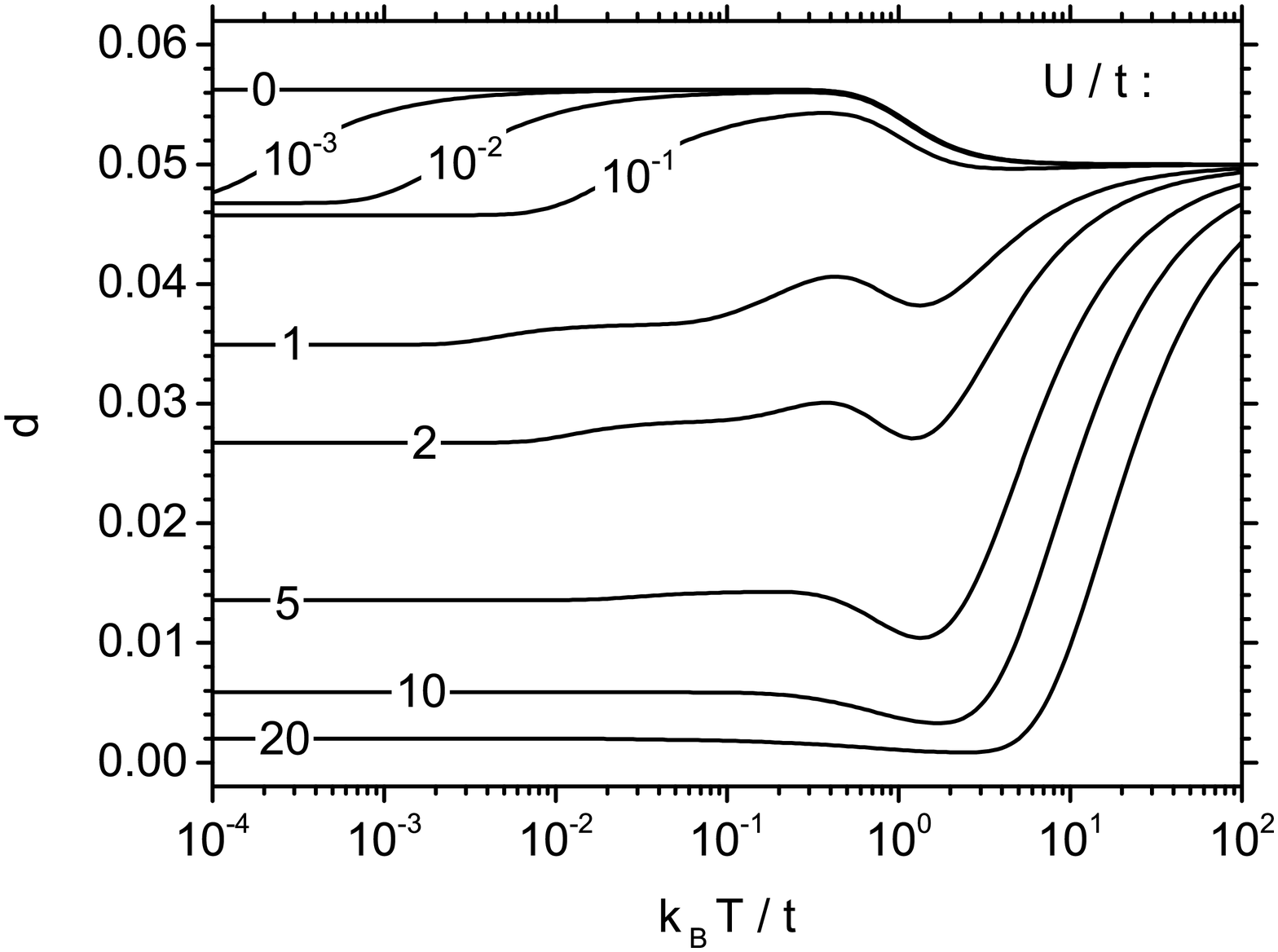}

\vspace{2mm}

\caption{The dependence of the average double occupancy per site on the normalized temperature for various values of normalized Coulombic interactions energy.\label{rys7}}
\end{center}
\end{figure}

As both specific heat and magnetic susceptibility exhibit the presence of pronounced maxima, it is interesting to investigate their behaviour as a function of Coulombic interactions energy as well as to explain their origin. 

The temperature values $T_{\max}$ at which such quantities as the magnetic susceptibility and the specific heat of the studied system reach their local maxima are plotted in Fig.~\ref{rys5} as a function of reduced Coulombic on-site energy $U/t$. Let us remind that, in general, susceptibility exhibits a single peak, while specific heat develops two distinct maxima. Therefore, in the main panel of Fig.~\ref{rys5}, only the position of low-temperature maximum of the specific heat is shown, while the inset presents the evolution of the position of high-temperature maximum. It can be noticed that the characteristic temperature of local maximum both for $\chi_{T}$ and for $C_h$ increases (starting from the zero value) until approximately $U/t\simeq 9$. If Coulombic interactions become stronger, the maxima shift back towards lower temperatures. The characteristic temperature at which the specific heat reaches maximum is at least twice lower than the corresponding temperature for magnetic susceptibility. The high-temperature maximum of specific hear shifts monotonically towards lower temperatures when $U/t$ increases, however, this evolution is quite rapid for weaker $U/t$ and then, for $U/t\gtrsim 9$ (where point of inflection is reached) it becomes significantly slower. 

In order to analyse the behaviour of the mentioned thermodynamic quantities in details, it can be useful to consider only the ground state and the first excited state. For the studied range of parameters ($0\leq U / t \leq 50$), the system of interest has a doubly degenerate ground state with total spin $S=0$. The first excited state is separated in energy from the ground state by the energy gap equal to $\Delta$ and has a degeneracy of nine, including three states with $S=-1$, three states with $S=0$ and three states with $S=1$. The normalized gap value $\Delta/t$ is plotted as a function of the ratio $U/t$ in Fig.~\ref{rys5} with dashed line (note the right vertical axis). It increases with increasing $U/t$ reaching a maximum value at $U/t\simeq 9.19$ and then a decrease is noticed. Such a knowledge about the energy spectrum allows us to analyse the behaviour of the selected thermodynamic quantities basing on a two-level system. 

According to fluctuation-dissipation theorem, the specific heat under constant magnetic field for the system with energy spectrum limited to the mentioned two states can expressed as:

\begin{equation}
C_{h}=k_{\rm B}\beta^2 \frac{9\Delta^2e^{-\beta\Delta}}{2+9e^{-\beta\Delta}}\left(1-\frac{9e^{-\beta\Delta}}{2+9e^{-\beta\Delta}}\right).
\label{eq4}
\end{equation}
This function exhibits a Schottky anomaly in a form of a broad maximum at $k_{\rm B}T_{\max}/\Delta \simeq 0.3264$.

Moreover, the magnetic susceptibility can be expressed under analogous assumptions as:
\begin{equation}
\chi_{T}=\frac{6\beta e^{-\beta\Delta}}{2+9e^{-\beta\Delta}},
\label{eq5}
\end{equation}
with a Schottky maximum at somehow higher temperature, equal to $k_{\rm B}T_{\max}/\Delta \simeq 0.5656$. 

The values of magnetic susceptibility and specific heat predicted by the equations (\ref{eq4}) and (\ref{eq5}) based on the calculated energy gap $\Delta$ are shown in Fig.~\ref{rys5} using empty and filled circles, respectively. It can be noticed, that for specific heat, a perfect agreement occurs between the results of full exact diagonalization calculations and the model involving only two states (ground state and the first excited state). Therefore, the origin of the low-temperature peak in specific heat can be explained in terms of a Schottky anomaly. In addition, the values for magnetic susceptibility also show a good agreement with the predictions of the two-level model, except at the highest studied values of $U/t$, i.e. for strongest Coulombic interactions.  Therefore, it can be concluded that the magnetic susceptibility peak and the low-temperature peak in specific heat possess the character of Schottky anomalies and they share the variability trend  with the energy gap $\Delta$ as a function of $U/t$. Moreover, they can be accurately described with a model involving only two states lowest in energy and arise only in the presence of Coulombic interactions.   On the other hand, the high-temperature maximum in specific heat is already present for a pure tight-binding model with $U/t=0$ and is unrelated to Coulombic interactions and only moderately sensitive to its occurrence.

The presented considerations show that the observed low-temperature maxima in magnetic susceptibility and specific heat are of Schottky anomaly origin, whereas the system of interest remains paramagnetic in the studied range of $U/t$. It can be noted that the similar plots in the Ref.~\cite{Zwicker} concerning specific heat for the case of 7 electrons show the effect of transition to the state with nonzero total spin over certain critical Coulombic interactions energy (which effect is absent in our case in the studied range).

In order to complete the characterization of the magnetic properties of the cubic cluster in the paramagnetic range, we illustrate the temperature dependence of the spin-spin correlations in Fig.~\ref{rys6}. In this plot two cases are presented - the absence of Coulombic interactions ($U/t=0$) and the presence of them, for a moderate value of $U/t=5$. As it can be noticed, the correlations between nearest-neighbour and second-neighbour spins are antiferromagnetic in character in both cases shown. On the contrary, the correlations betweens spins being third neighbours possess ferromagnetic character. In the absence of Coulombic interactions, the correlations exhibit almost flat temperature dependences unless the temperature is high, without any low-temperature features. Introducing the Coulombic interactions enhances very significantly the magnitude of correlations preserving their signs. Moreover, it leads to the appearance of low-temperature features, as the magnitude of correlations becomes more temperature dependent in some low-temperature range. This behaviour is much more pronounced for antiferromagnetic correlations, i.e. those between nearest-neighbours and second neighbours, whereas the ferromagnetic correlations between third neighbours remains less sensitive.

Last, but not least, we present the results of calculations of average double occupancy per site. The dependence of $d$ on the temperature for various values of Coulombic interactions energy is shown in Fig.~\ref{rys7}. It is remarkable that the low-temperature values of double occupancy strongly decrease when $U/t$ increases. Such a tendency of lowering the double occupancy with the increase of the Coulombic (repulsive) interactions energy is also visible for example in the results of Ref.~\cite{White1989} for two-dimensional Hubbard model. On the contrary, at high temperature the double occupancy increases (and tends to a common limit of $d=0.05$ at $T\to\infty$). Since the double occupancy of a site implies that a pair of electrons is in singlet state, with opposite spins, the increase in double occupancy is connected with a decrease in average squared magnetization per site (note that in the studied range the system is paramagnetic and the magnetizations themselves are equal to zero). Therefore, the temperature reduces the tendency to form magnetic moments (but increasing Coulombic interactions acts in an opposite direction). Ground-state value of double occupancy for $U/t=0$ is equal to $d=9/160=0.05625$, but the limiting value for $U/t\to 0$ is $d\simeq 0.04575$.

\section{Final remarks}

In the paper we discussed the thermodynamic properties of a cubic cluster described by Hubbard model at quarter-filling. The exact numerical diagonalization enabled obtaining the artefact-free solution. The studied cluster indicated paramagnetic properties with no magnetization in the investigated range of Coulombic interaction energies. The double-peak structure of specific heat was found, with a low-temperature maximum well described by a Schottky anomaly model, involving only the ground state and the first excited state. The same model was applied to explain the origin of a single maximum in magnetic susceptibility. Also the behaviour of spin-spin correlations and double occupancy was analysed. The obtained exact numerical results may encourage further studies of clusters with other geometries and numbers of electrons; moreover, also the influence of other factors, such as external fields, can be worthy of investigation.

\subsection*{Acknowledgments} This work has been supported by Polish Ministry of Science and Higher Education on a special purpose grant to fund the research and development activities and tasks associated with them, serving the development of young scientists and doctoral students.


\begin{thebibliography}{99}

\bibitem{Zukovic2014}
M.~\v{Z}ukovi\v{c}, A.~Bob\'ak, \href{http://dx.doi.org/10.1016/j.physleta.2014.04.063}{Phys. Lett. A \textbf{378} (2014) 1773-1779.}

\bibitem{Strecka2014}
J.~Stre\v{c}ka, J.~Ci\v{s}\'arov\'a, \href{http://dx.doi.org/10.12693/APhysPolA.126.26}{Acta Physica Polonica A \textbf{126} (2014) 26-27.}

\bibitem{Zukovic2015}
M.~\v{Z}ukovi\v{c}, \href{http://dx.doi.org/10.1016/j.jmmm.2014.08.017}{J. Magn. Magn. Mater. \textbf{374} (2015) 22-35.}



\bibitem{Strecka2015}
J.~Stre\v{c}ka, K.~Kar\'lov\'a, T.~Madaras, \href{http://dx.doi.org/10.1016/j.physb.2015.03.031}{Physica B \textbf{466-467} (2015) 76-85.}

\bibitem{Hubbard}
J.~Hubbard, \href{http://dx.doi.org/10.1098/rspa.1963.0204}{Proc. Roy. Soc. A \textbf{276} (1963) 238-257.}

\bibitem{Tasaki}
H.~Tasaki, \href{http://dx.doi.org/10.1088/0953-8984/10/20/004}{J. Phys.: Condens. Matter \textbf{10} (1998) 4353-4378.}

\bibitem{Spalek2015}
J.~Spa\l{}ek, \href{http://dx.doi.org/10.1080/14786435.2014.969352}{Philosophical Magazine \textbf{95} (2015) 661-681.}




\bibitem{Wojtczak1}
S.~Szczeniowski, L.~Wojtczak, \href{http://dx.doi.org/10.1051/jphyscol:19711420}{J. Phys. Colloques \textbf{32} (1971) C1-1174 - C1-1176.}

\bibitem{Wojtczak2}
A.~Sukiennicki, L.~Wojtczak, \href{http://dx.doi.org/10.1016/0375-9601(72)90622-6}{Phys. Lett. A \textbf{41} (1972) 37-38.}

\bibitem{Wojtczak3}
S.~Szczeniowski, L.~Wojtczak, A.~Sukiennicki, \href{http://dx.doi.org/10.1116/1.1318415}{J. Vac. Sci. Technol. \textbf{10} (1973) 693-696.}

\bibitem{Feshke}
A.~Wei{\ss}e, H.~Fehske, \href{http://dx.doi.org/10.1007/978-3-540-74686-7_18}{\emph{Exact Diagonalization Techniques}}, in: \emph{Computational Many-Particle Physics}, ed. by H.~Fehske, R.~Schneider, and A.~Wei{\ss}e, Lecture Notes in Physics vol. 739, Springer Berlin, Heidelberg, 2008, pp. 529-544.

\bibitem{Callaway1986}
J.~Callaway, D.P.~Chen, Y.~Zhang, \href{http://dx.doi.org/10.1007/BF01442352}{Z. Phys. D: Atom Mol. Cl. \textbf{3} (1986) 91-96.}



\bibitem{Callaway1987}
J.~Callaway, D.P.~Chen, Y.~Zhang, \href{http://dx.doi.org/10.1103/PhysRevB.36.2084}{Phys. Rev. B \textbf{36} (1987) 2084-2091.}

\bibitem{Callaway1988}
J.~Callaway, \href{http://dx.doi.org/10.1016/0378-4363(88)90212-4}{Physica B+C \textbf{149} (1988) 17-21.}

\bibitem{Parola}
A.~Parola, S.~Sorella, S.~Baroni, M.~Parrinello, E.~Tosatti,\\ \href{http://dx.doi.org/10.1142/S0217979289001202 }{Int. J. Mod. Phys. B \textbf{3} (1989) 1865-1873.}

\bibitem{Tan1991}
L.~Tan, Q.~Li, J.~Callaway, \href{http://dx.doi.org/10.1103/PhysRevB.44.341}{Phys. Rev. B \textbf{44} (1991) 341-350.}

\bibitem{Macedo}
A.~M.~S.~Mac\^{e}do, M. D. Coutinhe-Filho, \href{http://www.sbfisica.org.br/bjp/download/v21/v21a12.pdf}{Revista Brasileira de F\'isica \textbf{21} (1991) 121-135.}


\bibitem{Lopez2009}
F.~L\'{o}pez-Ur\'{i}as, G.M.~Pastor, \href{http://dx.doi.org/10.1140/epjd/e2009-00009-9}{Eur. Phys. J. D \textbf{52} (2009) 159-162.}


\bibitem{Zwicker}
D.~Zwicker, \emph{Gleichgewichtsthermodynamik des Hubbard-Modells f\"ur einen kubischen Cluster}, Diploma Thesis, Technische Universit\"at Dresden (2009).

\bibitem{Zwicker2}
R.~Schumann, D.~Zwicker, \href{http://dx.doi.org/10.1002/andp.201010452}{Ann. der Phys. (Berlin) \textbf{522} (2010) 419–439.}










\bibitem{Mathematica}
Wolfram Research, Inc., Mathematica, Version 8.04, Champaign, IL (2015).


\bibitem{White1989}
S.~R.~White, D.~J.~Scalapino, R.~L.~Sugar, E.~Y.~Loh, J.~E.~Gubernatis, R.~T.~Scalettar, \href{http://dx.doi.org/10.1103/PhysRevB.40.506}{Phys. Rev. B \textbf{40} (1989) 506-516.}


\end{thebibliography}
\end{document}